\documentclass[11pt]{article}
\usepackage{graphicx,amsmath,amssymb,amsfonts,hyperref}
\begin{document}

\title{\bf Preterm Birth Analysis Using Nonlinear Methods\\(a preliminary study)}
\author{Tijana Ivancevic, Lakhmi Jain, John Pattison and Alex Hariz\\ \emph{School of Electrical
and Information Engineering}\\ \emph{University of South
Australia}}\date{} \maketitle


\begin{abstract}
In this report we review modern nonlinearity methods that can be
used in the preterm birth analysis. The nonlinear analysis of
uterine contraction signals can provide information regarding
physiological changes during the menstrual cycle and pregnancy.
This information can be used both for the preterm birth prediction
and the preterm labor control. \bigbreak

\end{abstract}

\section{Introduction}

Preterm birth is the single most important cause of perinatal
mortality in North America and Europe \cite{Berkowitz}. Developed
countries represent 20 \% of the population in the world, but only
12 \% of human births annually, while 98 \% of medical
publications are issued from these areas. Reproductive patterns in
the developed world, for the last three decades, are different
from elsewhere and during the first 70 years of the 20th century.
A major difference is in the number of children in families but
also, and mainly, in the ages at first pregnancies in primiparae
(approaching now 30 years in many developed countries)
\cite{Dekker1}. A British study \cite{Rush} has estimated that
preterm birth accounts for 85 percent of early neonatal deaths
that are not caused by lethal congenital malformations. In the
United States, the smallest of the preterm infants—those weighing
less than 750 g have been found to account for 41 percent of early
neonatal deaths and 25 percent of infant deaths \cite{Overpeck}.
Preterm birth is also a major determinant of neonatal and infant
morbidity, including neuro-developmental handicaps, chronic
respiratory problems, intraventricular hemorrhage, infection,
retrolental fibroplasia, and necrotizing enterocolitis
\cite{Inst,Arias}. While there are inconsistent results with
regard to long-term somatic growth among preterm infants
\cite{Kitchen,Ross}, the risk of neurologic and developmental
impairment during childhood is substantially elevated for the
smallest survivors \cite{Escobar,Veen}. In addition, the neonatal
and long-term health care costs of preterm infants impose a
considerable economic burden both on individual families and
nationally \cite{Inst}.

While the frequency of births of low birth weight (less then 2,500
g) infants declined somewhat in the United States between 1970 and
1980, this decline appears to have occurred primarily among
full-term as opposed to preterm low birth weight infants \cite{Kessel}.
Furthermore, although the infant mortality rate has decreased
substantially since 1965, improvement in infant survival has been
attributed principally to reduced birth weight-specific mortality
rather than to changes in the birth weight or gestational age
distribution \cite{Kleinman}. Despite the reduction in infant mortality, the
rate in the United States remains considerably higher than the
rates in many other industrialized countries. It is unlikely that
there will be further substantial improvement in infant survival
in the United States unless a reduction in births of preterm low
birth weight infants can be accomplished \cite{Berkowitz}.

Traditionally, \emph{prematurity} was defined as a birth weight less
than or equal to 2,500 g. However, studies performed during the
1960s and 1970s revealed that this definition encompasses three
distinct types of infants: those who are small because they were
born too early, those who are small because their growth was
retarded in utero, and those who are small because they were both
premature and growth-retarded. The label \emph{prematurity} has now
been replaced by the terms \emph{low birth weight} and \emph{preterm}.
According to current World Health Organization nomenclature \cite{WHT1},
low birth weight characterizes an infant who weighs less than
2,500 g (5 pounds and 8 ounces) at birth, and preterm refers to a
birth that occurs at a gestational age of less than 37 completed
weeks (less then 259 days).

A cutoff of 37 weeks for defining preterm birth, albeit arbitrary,
is now well established, but some studies have used cutoffs of 36
or 38 weeks of gestation. In addition, some investigators have
only included preterm infants of a particular weight, such as
those weighing less than 2,500 g. Further subdivisions have also
been used in the recent literature, such as \emph{very low birth
weight} to describe an infant weighing less than 1,500 g or 1,000
g at birth, and \emph{very preterm} for gestational ages variously
defined as less than 32, 33, or 34 weeks. Use of these lower
cutoffs should be encouraged, since it will permit identification
of risk factors that have the greatest impact on neonatal
mortality \cite{Berkowitz}.

For detecting contractions in \emph{uterine electromyography}
(EMG), fast numerical algorithms have been developed
\cite{Radharkrishnan}. Recently, the analysis of \emph{uterine
contraction signals} has provided information regarding
physiological changes during the menstrual cycle and pregnancy
\cite{Oczeretko1}. The authors have presented the
\emph{cross-correlation} and the \emph{wavelet cross-correlation}
methods to assess synchronization between contractions in
different topographic regions of the uterus. It is important to
identify time delays between uterine contractions, which may be of
potential diagnostic significance in various pathologies. The
cross-correlation was computed in a moving window with a width
corresponding to approximately two or three contractions. As a
result, the running cross-correlation function was obtained. The
propagation \% parameter assessed from this function allows
quantitative description of synchronization in bivariate time
series. In general, the uterine contraction signals are very
complicated. \emph{Wavelet transforms} provide insight into the
structure of the time series at various frequencies (scales). To
show the changes of the propagation \% parameter along scales, a
wavelet running cross-correlation was used. At first, the
continuous wavelet transforms as the uterine contraction signals
were received and afterwards, a running cross-correlation analysis
was conducted for each pair of transformed time series. The
findings of \cite{Oczeretko1} show that running functions are very
useful in the analysis of uterine contractions.

\emph{Fractal analysis} serves another example for advanced data
analysis to be utilized in the studies on uterine contractile
activity. Fractals, a relatively new analytical concept of the
last few decades, have been successfully applied in many areas of
science and technology. \cite{Oczeretko0} performed a comparative
study of ten methods of fractal analysis of signals of
intrauterine pressure, and found significant differences between
uterine contractions in healthy volunteers and women with primary
dysmenorrhoea. There was a correlation between the adjacent
elements of the investigated signals. Consequently, the values of
fractal dimension can be objective measures for classifications of
uterine contraction signals and may be used in studies on preterm
labor.

\emph{Nonlinear dynamics} is one more example of advanced data
analysis that could be implemented for uterine contractility
signals \cite{Pierzynski}. In recent years the physiological
signals obtained from the complex systems like the brain or the
heart have been investigated for possible deterministic chaotic
behavior. The human uterus is undoubtedly a complex system. Smooth
muscles comprising the myometrium interact in a complex manner.
The techniques of surrogate data analysis have been used to
testing for nonlinearity in the uterine contraction signals. The
results showed that the spontaneous uterine contractions are
considered to contain nonlinear features \cite{Oczeretko} which
indicated that nonlinear dynamics may increase the accessibility
of data for the assessment of biology of uterus, and consequently,
the nature of preterm labor.

\section{Techniques of Nonlinear Dynamics and Complex Data Analysis}

In this section we review the \emph{concept of nonlinearity} (as
contrasted with the standard textbook linear algebra, analysis,
statistics, etc.), together with its geometrical picture, and its
extreme dynamics of chaotic behavior.

\subsection{Nonlinear Dynamics}

Recall that \emph{nonlinear dynamics} is a modern language to talk
about dynamical systems. It is a general theory of systems arising
in physics, engineering, chemistry, biology, psychology, sociology
and economics. Nonlinear dynamics has two extremes: at one end it
is classical linear dynamics, fully predictable and controllable,
as usually explained in engineering textbooks. On the other end,
it is \emph{chaotic dynamics}, or popularly, the ``chaos
theory."\footnote{All chaotic systems are nonlinear, but not all
nonlinear systems are chaotic. Thus, chaotic dynamics is an
extreme subset of nonlinear dynamics.} Nonlinear dynamics includes
both of these extremes, as well as all other natural and
artificial dynamics. Its main keywords are (see, e.g.,
\cite{StrAttr}):
\begin{itemize}
\item \emph{Dynamical system: }A part of the world which can be
seen as a self--contained entity with some temporal behavior. In
nonlinear dynamics, speaking about a dynamical system usually
means to speak about an abstract mathematical system which is a
model for such an entity. Mathematically, a
dynamical system is defined by its \textit{state} and by its \textit{dynamics%
}. A pendulum is an example for a dynamical system.

\item \emph{State of a system: }A number or a vector (i.e., a list
of numbers) defining the state of the dynamical system uniquely.
For the free (un--driven) pendulum, the state is uniquely defined
by the angle $\theta $\ and the angular velocity
$\dot{\theta}=d\theta /dt$. In the case of driving, the driving
phase $\phi $\ is also needed because the pendulum becomes a
non--autonomous system. In spatially extended systems, the state
is often a \textit{field} (a scalar--field or a vector--field).
Mathematically spoken, fields are functions with space coordinates
as independent variables. The velocity vector--field of a fluid is
a well--known example.

\item \emph{Phase space: }All possible states of the system. Each
point in the phase--space corresponds to a unique state. In the case of the free pendulum, the
phase--space has 2D whereas for driven pendulum it has 3D. The
dimension of the phase--space is infinite in cases where the
system state is defined by a field.

\item \emph{Dynamics, or equation of motion: }The causal relation
between the present state and the next state in the future. It is
a deterministic rule which tells us what happens in the next time
step. In the case of a continuous time, the time step is
infinitesimally small. Thus, the equation of motion is an ordinary
differential equation (ODE) (or a system of ODEs):
\begin{equation*}
\dot{x}=f(x),
\end{equation*}%
where $x$ is the state and $t$ is the time variable (overdot is
the time derivative -- as always). An example is the equation of
motion of an un--driven and un--damped pendulum. In the case of a
discrete time, the time steps are nonzero and the dynamics is a
map:
\begin{equation*}
x_{n+1}=f(x_{n}),
\end{equation*}%
with the discrete time $n$. Note, that the corresponding physical
time points $t_{n}$ do not necessarily occur equidistantly. Only
the order has to be the same. That is,
\begin{equation*}
n<m\qquad \Longrightarrow \qquad t_{n}<t_{m}.
\end{equation*}%
The dynamics is \emph{linear} if the causal relation between the
present state and the next state is linear. Otherwise it is
\emph{nonlinear}. If we have the case in which the next state is
not uniquely defined by the present one, this is generally an
indication that the \emph{phase--space is not complete}. Thus,
there are important variables determining the state which had been
forgotten. This is a crucial point while modelling a real--life
systems. Beside this, there are two important classes of systems
where the
phase--space is incomplete: the \emph{non--autonomuous and stochastic systems}%
. A non--autonomous system has an equation of motion which depends
explicitly on time. Thus, the dynamical rule governing the next
state not only depends on the present state but also at the time
it applies. A driven pendulum is a classical example of a
\textit{non--autonomuous system}. Fortunately, there is an easy
way to make the phase--space complete: we simply include the time
into the definition of the state. Mathematically, this is done by
introducing a new state variable: $t$. Its dynamics reads
\begin{equation*}
\dot{t}=1,\qquad \text{or}\qquad t_{n+1}=t_{n},
\end{equation*}%
depending on whether time is continuous or discrete. For the
periodically driven pendula, it is also natural to take the
driving phase as the new state variable. Its equation of motion
reads
\begin{equation*}
\dot{\theta}=2\pi w,
\end{equation*}%
where $w$ is the driving frequency (so that the angular driving
frequency is $2\pi w$). On the other hand, in a \textit{stochastic
system}, the number and the nature of the variables necessary to
complete the phase--space is usually unknown. Therefore, the next
state can not be deduced from the present one. The deterministic
rule is replaced by a stochastic one. Instead of the next state,
it gives only the probabilities of all points in the phase--space
to be the next state.

\item \emph{Orbit or trajectory: }A solution of the equation of
motion. In the case of continuous time, it is a curve in
phase--space parameterized by the time variable. For a discrete
system it is an ordered set of points in the phase--space.

\item \emph{Phase Flow: }The mapping (or, map) of the whole
phase--space of a continuous dynamical system onto itself for a
given time step $t$. If $t$ is an infinitesimal time step $dt$,
the flow is just given by the right--hand side of the equation of
motion (i.e., $f$). In general, the flow for a finite time step is
not known analytically because this would be equivalent to have a
solution of the equation of motion.
\end{itemize}

\begin{figure}[htb]
\centerline{\includegraphics[width=4cm]{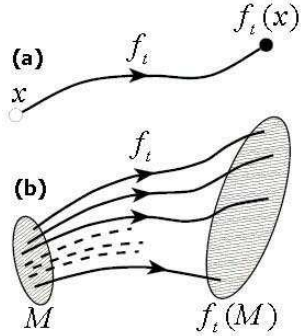}} \caption{Action
of the \emph{phase--flow} $f_t$ in the phase--space manifold $M$:
(a) Trajectory of a single initial point $x(t)\in M$ , (b)
Transporting the whole manifold $M$ (see \cite{StrAttr}).}
\label{CvitFlow}
\end{figure}

More precisely, a mathematical term \textit{dynamical system}
geometrically represents a \textit{vector--field} in the system's
phase--space manifold $M$ (see \cite{GaneshSprBig}), which upon
\textit{integration} (governed by the celebrated \textit{existence
\& uniqueness theorems for ordinary Differential Equations})
defines a \textit{phase--flow} in $M$ (see Figure \ref{CvitFlow}).
This phase--flow $f_t\in M$, describing the complete behavior of a
dynamical system at every time instant, can be either linear,
nonlinear or chaotic.

Before the advent of fast computers, solving a dynamical system
required sophisticated mathematical techniques and could only be
accomplished for a small class of linear dynamical systems.
Numerical methods executed on computers have simplified the task
of determining the orbits of a dynamical system.

For simple dynamical systems, knowing the trajectory is often
sufficient, but most dynamical systems are too complicated to be
understood in terms of individual trajectories. The difficulties
arise because:

1. The systems studied may only be known approximately--the
parameters of the system may not be known precisely or terms may
be missing from the equations. The approximations used bring into
question the validity or relevance of numerical solutions. To
address these questions several notions of stability have been
introduced in the study of dynamical systems, such as
\textit{Lyapunov stability} or \textit{structural stability}. The
stability of the dynamical system implies that there is a class of
models or initial conditions for which the trajectories would be
equivalent. The operation for comparing orbits to establish their
equivalence changes with the different notions of stability.

2. The type of trajectory may be more important than one
particular trajectory. Some trajectories may be periodic, whereas
others may wander through many different states of the system.
Applications often require enumerating these classes or
maintaining the system within one class. Classifying all possible
trajectories has led to the qualitative study of dynamical
systems, that is, properties that do not change under coordinate
changes. Linear dynamical systems and systems that have two
numbers describing a state are examples of dynamical systems where
the possible classes of orbits are understood.

3. The behavior of trajectories as a function of a parameter may
be what is needed for an application. As a parameter is varied,
the dynamical systems may have \textit{bifurcation point}\emph{s}
where the qualitative behavior of the dynamical system changes.
For example, it may go from having only periodic motions to
apparently erratic behavior, as in the transition to
\textit{turbulence} of a fluid.

4. The trajectories of the system may appear erratic, as if
random. In these cases it may be necessary to compute averages
using one very long trajectory or many different trajectories. The
averages are well defined for \emph{ergodic
systems} and a more detailed understanding has been worked out for \textit{%
hyperbolic systems}. Understanding the probabilistic aspects of
dynamical systems has helped establish the foundations of
statistical mechanics and of chaos.

\subsection{Geometrical View on Nonlinear Dynamics}

Geometrically speaking (see \cite{GaneshSprBig}), \textit{linear
systems} (i.e., systems defined by linear differential equations)
live in \textit{Euclidean spaces} $\mathbb{R}^{N}$, where $N$ is
the dimension of the system. For their analysis the tools of
linear algebra and calculus are used. The basic measure here is the \textit{%
global Euclidean metric form} (or, Euclidean distance function)
function on the system coordinates, defined by
\[
S=\sqrt{\sum_{i=1}^{N}\left( x_{i}-y_{i}\right) ^{2}},\qquad
\text{or\qquad }S^{2}=\sum_{i=1}^{N}\left( x_{i}-y_{i}\right)
^{2}.
\]

The more realistic \textit{nonlinear systems} (i.e., systems
defined by non-linear differential equations) can be considered as
local deformations
of the closest linear systems and they live in \textit{smooth manifolds} $%
{M}^{N}.$ For their analysis the tools of Riemannian
geometry are
used, with the \textit{local Riemannian metric form}, defined by%
\begin{equation}
ds^{2}=\sum_{i=1}^{N}\sum_{j=1}^{N}g_{ij}dx_{i}dx_{j},\qquad
\text{or\qquad }ds^{2}=g_{ij}dx^{i}dx^{j},  \label{Riemann}
\end{equation}%
where $g_{ij}$ is the \textit{Riemannian metric tensor}, $dx^{i}$
are differentials of the local coordinates, and the
\textit{Einstein's summation convention} (summing upon repeated
indices if one is superscript-contravariant and the other is
subscript-covariant) is in place. Besides giving the local
distances between the points on the smooth manifold
$\mathbb{M}^{N},$ the Riemannian metric form (\ref{Riemann})
defines the
system's kinetic energy%
\[
KE=\frac{1}{2}g_{ij}\dot{x}^{i}\dot{x}^{j},\qquad (\text{overdot
means time derivative}),
\]
giving the Euler-Lagrangian equations of motion in a geometrical form (for $s=t$) of \emph{geodesic equations}
\begin{equation}
\frac{d^{2}x^{i}}{ds^{2}}+\Gamma _{jk}^{i}\frac{dx^{j}}{ds}\frac{dx^{k}}{ds}%
=0,  \label{geod-mot}
\end{equation}%
where $\Gamma _{jk}^{i}$ are the so--called Christoffel symbols of
the affine Levi-Civita connection of the manifold ${M}^{N}$ (see, e.g., \cite{GaneshADG}).

In the geometrical framework, the (in)stability of the
trajectories is the (in)stability of the geodesics, and it is
completely determined by the curvature properties of the
underlying manifold according to the \textit{Jacobi equation} of
\emph{geodesic deviation}\\ \cite{GaneshADG}
\begin{equation}
\frac{D^{2}J^{i}}{ds^{2}}+R_{~jkm}^{i}\frac{dx^{j}}{ds}J^{k}\frac{dx^{m}}{ds}%
=0,  \label{eqJ}
\end{equation}%
whose solution $J$, usually called \textit{Jacobi variation
field}, locally measures the distance between nearby geodesics;
$D/ds$ stands for the \textit{covariant derivative} along a
geodesic and $R_{~jkm}^{i}$ are the components of the
\textit{Riemann curvature tensor}.

In the special case of the \textit{Eisenhart metric} \cite{Eisenhart} on an enlarged
configuration space--time ($\{x^{0}\equiv t,x^{1},\ldots ,x^{N}\}$,
the Jacobi equation (\ref{eqJ}) reduces to the \emph{tangent dynamics equation}
\cite{GaneshADG}
\begin{equation}
\frac{d^{2}J^{i}}{dt^{2}}+\left( \frac{\partial ^{2}V}{\partial
x^{i}\partial x^{k}}\right) _{x(t)}J^{k}=0.  \label{Tg-dyn}
\end{equation}%

\subsection{Extreme Nonlinearity: Chaotic Dynamics}

On the other hand, \emph{chaotic dynamics}, the most extreme case
of nonlinear dynamics, is highly sensitive to both initial
conditions (popularly referred to as the \emph{butterfly effect})
and system parameters. As a result of this sensitivity, which
manifests itself as an exponential growth of perturbations in the
initial conditions, the behavior of chaotic systems appears to be
random. This happens even though these systems are deterministic,
meaning that their future dynamics are fully defined by their
initial conditions, with no random elements involved. This
behavior is known as \emph{deterministic chaos} (see, e.g.,
\cite{StrAttr}). Chaotic behavior has been observed in the
laboratory in a variety of systems including electrical circuits,
lasers, oscillating chemical reactions, fluid dynamics, and
mechanical and magneto-mechanical devices.

Chaotic behavior is surprisingly complex and may arise in a simple
noise-free system with few degrees of freedom. Chaos is globally
stable and locally unstable. A chaotic system neither reaches a
certain steady state nor repeats itself periodically. Thus, it is
a pure disordered process since no points or patterns of points
ever recur. However, its activity remains constrained within a
certain boundary, which may suggest a new kind of order. Chaotic
behavior resembles random noise, but is predictable in the
short-term. This short-term predictability is useful in various
domains ranging from weather forecasting to economic forecasting
\cite{Hilborn}.

The most common \emph{route to chaos} is a sequence of
period-doubling bifurcations, caused by particular values of a
nonlinear system parameter. The cases of most interest arise when
the chaotic behavior takes place on an \emph{attractor}, since
then a large set of initial conditions will lead to orbits that
converge to this chaotic region. In particular, the so--called
\emph{strange attractors} typically have a fractal structure. In
all chaotic systems all particle trajectories diverge
exponentially from one another, with a positive \emph{Lyapunov
exponent}. Besides, chaotic behavior is related to the
\emph{non-equilibrium phase transitions}\footnote{Phase
transitions (PT) are phenomena which bring about
\emph{qualitative} physical changes at the macroscopic level in
presence of the same microscopic forces acting among the
constituents of a system (e.g.,
$Solid~\leftrightarrows~Liquid~\leftrightarrows~Gas~\leftrightarrows~Plasma$).
Their mathematical description requires to translate into
\emph{quantitative} terms the mentioned qualitative changes. The
standard way of doing this is to consider how the values of
thermodynamic observables, obtained in laboratory experiments,
vary with temperature, or volume, or an external field, and then
to associate the experimentally observed discontinuities at a PT
to the appearance of some kind of singularity entailing a loss of
analyticity \cite{FP04}. The so--called non-equilibrium phase
transitions have been elaborated in the realm of
\emph{synergetics} \cite{Haken,Haken2,NatBiodyn} as a ubiquitous
route to self-organization in complex systems of various nature
(e.g., in brain's function \cite{HakenBrain}).} caused by system's
\emph{topology change}\footnote{\emph{Topology change} means loss
of the system's
diffeomorphicity. Namely, the so--called \textit{%
topological theorem} \cite{FP04} says that non--analyticity is the
``shadow" of a more fundamental
phenomenon occurring in the system's configuration space: a topology change
within the family of equipotential hyper-surfaces
\[
\Sigma _{v}=\{(x_{1},\dots ,x_{N})\in \mathbb{R}^{N}|\
V(x_{1},\dots ,x_{N})=v\},
\]%
where $V=V(x)$ is the microscopic interaction potential expressed in
the coordinates $x_i$. The
\emph{largest Lyapunov exponent} $\lambda _{1}$ (see subsection \ref{Lj} below)
is computed by solving the above \textit{tangent dynamics equation} (\ref{Tg-dyn})
so that we get \cite{FP04}
\[
\lambda _{1}=\lim_{t\rightarrow \infty }1/2t\log (\Sigma _{i=1}^{N}[\dot{J}%
_{i}^{2}(t)+J _{i}^{2}(t)]/\Sigma
_{i=1}^{N}[\dot{J}_{i}^{2}(0)+J _{i}^{2}(0)]).
\]} \cite{CCC97,FCS99,FPS00}.

Techniques have emerged to harness chaos to the benefit of humans.
The chaotic behavior of a system may be artificially weakened or
suppressed if it is undesirable. This concept is known as
\emph{control of chaos}. The first and the most important method
of chaos control is the so--called OGY--method, developed by
\cite{OGY}. However, in recent years, a non-traditional concept of
anti-control of chaos has emerged. Here, the non-chaotic dynamical
system is transformed into a chaotic one by small controlled
perturbation so that useful properties of a chaotic system can be
exploited. This non-traditional concept has found applications in
time and energy critical control applications such as navigation
in a multi-body planetary system, fluid mixing and to secure
information processing \cite{ChOrder}.

\subsubsection{Lorenz Attractor}

\begin{figure}[tbh]
\centerline{\includegraphics[height=4cm]{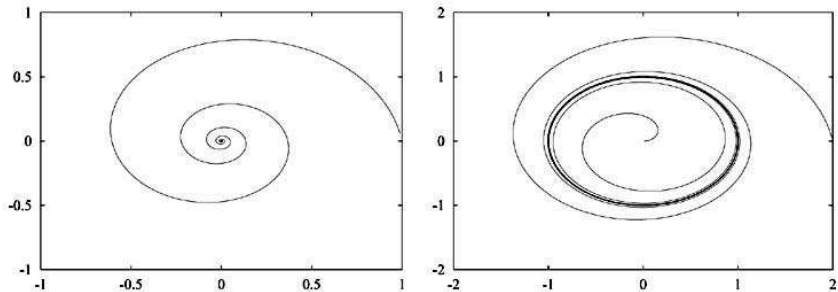}} \caption{Examples
of regular attractors: fixed--point (left) and limit cycle
(right). Note that limit cycles exist only in nonlinear dynamics.}
\label{fig:regular}
\end{figure}

Recall that a dynamical system may be defined as a deterministic
rule for the time evolution of state observables. Well known
examples are \textit{ODEs} in which time is continuous
\cite{StrAttr}
\begin{equation}
\mathbf{\dot{x}}(t)=\mathbf{f}(\mathbf{x}(t)),\;\;\;\;\;(\mathbf{x},\mathbf{f%
}\in \mathbb{R}^{n});  \label{eq:ode}
\end{equation}%
and \textit{iterative maps} in which time is discrete:
\begin{equation}
\mathbf{x}(t+1)=\mathbf{g}(\mathbf{x}(t)),\;\;\;\;\;(\mathbf{x},\mathbf{g}%
\in \mathbb{R}^{n}).  \label{eq:maps}
\end{equation}%
In the case of maps, the evolution law is straightforward: from $\mathbf{x}%
(0)$ one computes $\mathbf{x}(1)$, and then $\mathbf{x}(2)$ and so
on. For ODE's, under rather general assumptions on $\mathbf{f}$,
from an initial
condition $\mathbf{x}(0)$ one has a unique trajectory $\mathbf{x}(t)$ for $%
t>0$ \cite{Ott93}. Examples of regular behaviors (e.g., stable
fixed--points, limit cycles) are well known, see
Figure~\ref{fig:regular}.

A rather natural question is the possible existence of less
regular behaviors i.e., different from stable fixed--points,
periodic or quasi-periodic motion.

\begin{figure}[htb]
\centerline{\includegraphics[width=7cm]{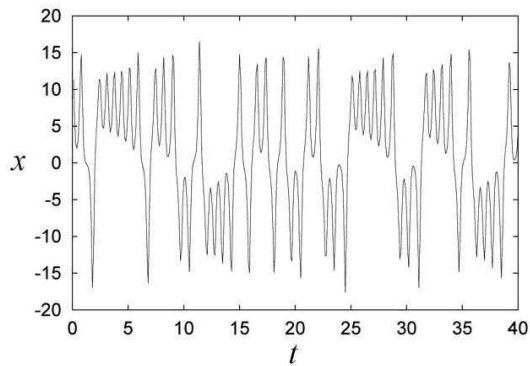}}
\caption{Example of an aperiodic signal: the $x$ variable of the Lorenz system (%
\ref{eq:lorenz}) as function of time $t$, for $r = 28$.}
\label{fig:aperiod}
\end{figure}

After the seminal works of Poincar\'{e}, Lorenz, Smale, May, and
H\'enon (to cite only the most eminent ones) it is now well
established that the so called \emph{chaotic behavior} is
ubiquitous. As a relevant system, originated in the geophysical
context, we mention the celebrated \textit{Lorenz system}
\cite{Lorenz,Sp}
\begin{eqnarray}
\dot{x} &=&-\sigma (x-y)  \notag \\
\dot{y} &=&-xz+rx-y  \label{eq:lorenz} \\
\dot{z} &=&xy-bz  \notag
\end{eqnarray}%
This system is related to the \textit{Rayleigh--B\'enard
convection} under very crude approximations. The quantity $x$ is
proportional the circulatory fluid particle velocity; the
quantities $y$ and $z$ are related to the temperature profile;
$\sigma $, $b$ and $r$ are dimensionless parameters. Lorenz
studied the case with $\sigma =10$ and $b=8/3$ at varying $r$
(which is proportional to the Rayleigh number). It is easy to see
by linear analysis that the fixed--point $(0,0,0)$ is stable for
$r<1$. For $r>1$ it becomes unstable and two new fixed--points
appear
\begin{equation}
C_{+,-}=(\pm \sqrt{b(r-1)},\pm \sqrt{b(r-1)},r-1),
\label{eq:fixedpts}
\end{equation}%
these are stable for $r<r_{c}=24.74$. A nontrivial behavior, i.e.,
non periodic, is present for $r>r_{c}$, as is shown in
Figure~\ref{fig:aperiod}.

In this `strange', chaotic regime one has the so called sensitive
dependence
on initial conditions. Consider two trajectories, $\mathbf{x}(t)$ and $%
\mathbf{x}^{\prime}(t)$, initially very close and denote with $\Delta(t)=||%
\mathbf{x}^{\prime}(t)-\mathbf{x}(t)||$ their separation. Chaotic
behavior means that if $\Delta(0) \to 0$, then as $t \to \infty$
one has $\Delta(t) \sim \Delta(0) \exp{\lambda_1 t}$, with
$\lambda_1 > 0$ \cite{BLV01}.

Let us notice that, because of its chaotic behavior and its
dissipative nature, i.e.,
\begin{equation}
{\frac{\partial \dot x }{\partial x}} + {\frac{\partial \dot y
}{\partial y}} + {\frac{\partial \dot z }{\partial z}} < 0,
\label{eq:divlorenz}
\end{equation}
the attractor of the Lorenz system cannot be a smooth surface.
Indeed the attractor has a self--similar structure with a fractal
dimension between $2$ and $3$. The Lorenz model (which had an
important historical relevance in the development of chaos theory)
is now considered a paradigmatic example of a chaotic system.

\subsubsection{Lyapunov Exponents}\label{Lj}

The sensitive dependence on the initial conditions can be
formalized in order to give it a quantitative characterization.
The main growth rate of trajectory separation is measured by the
first (or maximum) \textit{Lyapunov exponent}, defined as (see,
e.g., \cite{BLV01})
\begin{equation}
\lambda _{1}=\lim_{t\rightarrow \infty }\lim_{\Delta (0)\rightarrow 0}{\frac{%
1}{t}}\ln {\frac{\Delta (t)}{\Delta (0)}},  \label{eq:lyap1}
\end{equation}%
As long as $\Delta (t)$ remains sufficiently small (i.e.,
infinitesimal,
strictly speaking), one can regard the separation as a tangent vector $%
\mathbf{z}(t)$ whose time evolution is
\begin{equation}
\dot{z}_{i}={\frac{\partial f_{i}}{\partial
x_{j}}}|_{\mathbf{x}(t)}\cdot z_{j},  \label{eq:tangent}
\end{equation}%
and, therefore,
\begin{equation}
\lambda _{1}=\lim_{t\rightarrow \infty }{\frac{1}{t}}\ln {\frac{||\mathbf{z}%
(t)||}{||\mathbf{z}(0)||}}.  \label{eq:lambdatang}
\end{equation}%
In principle, $\lambda _{1}$ may depend on the initial condition $\mathbf{x}%
(0)$, but this dependence disappears for ergodic systems. In
general there exist as many Lyapunov exponents, conventionally
written in decreasing order $\lambda _{1}\geq \lambda _{2}\geq
\lambda _{3}\geq ...$, as the independent coordinates of the
phase--space \cite{BGGS80}. Without entering the details, one can
define the sum of the first $k$ Lyapunov exponents as the growth
rate of an infinitesimal $k$D volume in the phase--space. In particular, $%
\lambda _{1}$ is the growth rate of material lines, $\lambda
_{1}+\lambda _{2}$ is the growth rate of $2D$ surfaces, and so on.
A numerical widely used efficient method is due to \cite{BGGS80}.

It must be observed that, after a transient, the growth rate of
any generic small perturbation (i.e., distance between two
initially close trajectories)
is measured by the first (maximum) Lyapunov exponent $\lambda_1$, and $%
\lambda_1 > 0$ means chaos. In such a case, the state of the
system is unpredictable on long times. Indeed, if we want to
predict the state with a certain tolerance $\Delta$ then our
forecast cannot be pushed over a certain
time interval $T_P$, called \emph{predictability time}, given by \cite{BLV01}%
:
\begin{equation}
T_P \sim {\frac{1 }{\lambda_1}} \ln {\frac{\Delta }{\Delta(0)}}.
\label{eq:predtime}
\end{equation}
The above relation shows that $T_P$ is basically determined by $1/\lambda_1$%
, seen its weak dependence on the ratio $\Delta/\Delta(0)$. To be
precise one must state that, for a series of reasons, relation
(\ref{eq:predtime}) is too simple to be of actual relevance
\cite{PredictCompl}.

\subsubsection{Kolmogorov--Sinai Entropy}

Deterministic chaotic systems, because of their irregular
behavior, have many aspects in common with stochastic processes.
The idea of using stochastic processes to mimic chaotic behavior,
therefore, is rather natural \cite{chirikov79,B84}. One of the
most relevant and successful approaches is symbolic dynamics
\cite{BS93}. For the sake of simplicity let us consider a discrete
time dynamical system. One can introduce a partition $\mathcal{A}$
of the phase--space formed by $N$ disjoint sets $A_{1},...,A_{N}$.
From any initial condition one has a trajectory
\begin{equation}
\mathbf{x}(0)\rightarrow
\mathbf{x}(1),\mathbf{x}(2),...,\mathbf{x}(n),... \label{eq:traj}
\end{equation}%
dependently on the partition element visited, the trajectory (\ref{eq:traj}%
), is associated to a symbolic sequence
\begin{equation}
\mathbf{x}(0)\rightarrow i_{1},i_{2},...,i_{n},...
\label{eq:symbol}
\end{equation}%
where $i_{n}$ ($n=1,2,...,N$) means that $\mathbf{x}(n)\in
A_{i_{n}} $ at the step $n$, for $n=1,2,...$. The coarse-grained
properties of chaotic trajectories are therefore studied through
the discrete time process (\ref{eq:symbol}).

An important characterization of symbolic dynamics is given by the \textit{%
Kolmogorov--Sinai entropy} (KS), defined as follows. Let $%
C_{n}=(i_{1},i_{2},...,i_{n})$ be a generic `word' of size $n$ and
$P(C_{n})$ its occurrence probability, the quantity \cite{BLV01}
\begin{equation}
H_{n}=\sup_{A}[-\sum_{C_{n}}P(C_{n})\ln P(C_{n})],
\label{eq:block}
\end{equation}%
is called \textit{block entropy} of the $n-$sequences, and it is
computed by taking the largest value over all possible partitions.
In the limit of infinitely long sequences, the asymptotic entropy
increment
\begin{equation}
h_{KS}=\lim_{n\rightarrow \infty }H_{n+1}-H_{n},  \label{eq:KS}
\end{equation}%
is the Kolmogorov--Sinai entropy. The difference $H_{n+1}-H_{n}$
has the intuitive meaning of average information gain supplied by
the $(n+1)-$th symbol, provided that the previous $n$ symbols are
known. KS--entropy has an
important connection with the positive Lyapunov exponents of the system \cite%
{Ott93}:
\begin{equation}
h_{KS}=\sum_{\lambda _{i}>0}\lambda _{i}.  \label{eq:pesin}
\end{equation}%
In particular, for low--dimensional chaotic systems for which only
one Lyapunov exponent is positive, one has $h_{KS}=\lambda _{1}$.

We observe that in (\ref{eq:block}) there is a technical
difficulty, i.e., taking the sup over all the possible partitions.
However, sometimes there exits a special partition, called
generating partition, for which one finds that $H_{n}$ coincides
with its superior bound. Unfortunately the generating partition is
often hard to find, even admitting that it exist. Nevertheless,
given a certain partition, chosen by physical intuition, the
statistical properties of the related symbol sequences can give
information on the dynamical system beneath. For example, if the
probability of observing a symbol (state) depends only by the
knowledge of the immediately preceding symbol, the symbolic
process becomes a \textit{Markov chain} (see \cite{GaneshSprBig})
and all the statistical properties are determined by the transition matrix elements $%
W_{ij}$ giving the probability of observing a transition
$i\rightarrow j$ in one time step. If the memory of the system
extends far beyond the time step between two consecutive symbols,
and the occurrence probability of a symbol depends on $k$
preceding steps, the process is called \textit{Markov process} of
order $k$ and, in principle, a $k$ rank tensor would be required
to describe the dynamical system with good accuracy. It is
possible to demonstrate that if $H_{n+1}-H_{n}=h_{KS}$ for $n\geq
k+1$, $k$ is the (minimum) order of the required Markov process.
It has to be pointed out, however, that to know the order of the
suitable Markov process we need is of no practical utility if
$k\gg 1$.

\subsection{Nonlinear Analysis of Time Series Data}

In this subsection, following \cite{Sanjeev}, we apply two basic
techniques of nonlinear dynamics to the \emph{heart inter-beat
interval} (IBI) time series data. The volunteer participant was a
twenty year-old male pilot with 400 hours total flight experience
and a commercial pilot license. He flew a single-engine aircraft
in a fixed-base generic flight simulator. A Polar Heart-rate
Receiver/Transmitter was used to record his heart IBI. The task of
the participant in the flight segments M1 and M2 was to maintain
manually a set airspeed, altitude, and heading as closely as
possible. The only difference between the two segments, M1 and M2,
was the level of perceived risk of mid-air collision involved if
the participant was unable to execute the task adequately. First,
a sample participant's heart IBI data at segments M1 and M2 are
represented graphically and analyzed in terms of phase plots.
Next, the measure of approximate entropy is employed to quantify
the dynamics for each time series \cite{Sanjeev}.

\subsubsection{Phase Plots}

The qualitative dynamics of a system may be represented in a
graphical form. The Phase plot is the simplest graphical
representation of the temporal information contained in a time
series. It relates the current value of the time series to its
preceding value. However, the preceding value is expressed in such
a manner that it is approximately equal to the derivative of the
current value. The phase plot provides a spatial representation of
the evolving dynamics of a nonlinear system, giving some
information on how the process evolves over time \cite{Williams}.
\begin{figure}[htb]
\centerline{\includegraphics[width=14cm]{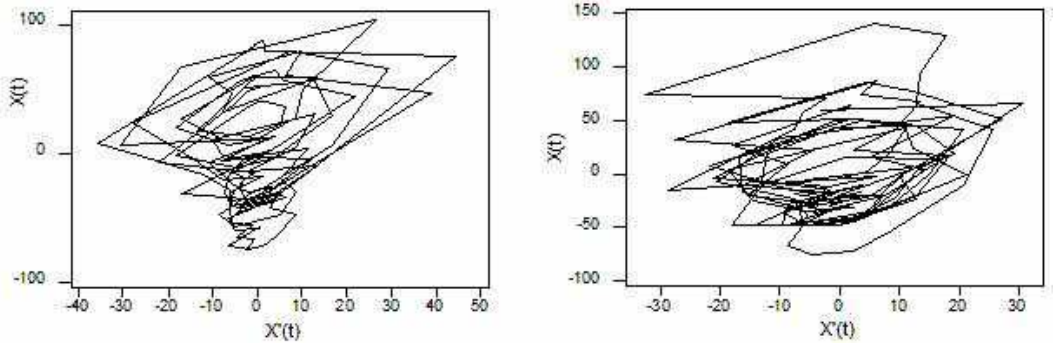}}
\caption{Two-dimensional phase plots of heart inter-beat interval
(IBI) for a pilot at the simulated flying segments M1 (left) and
M2 (right; adapted from \cite{Sanjeev}).} \label{Sanjeev1}
\end{figure}

Figure \ref{Sanjeev1} represents two-dimensional phase plots
employed to depict the graphical dynamics of a sample
participant's heart IBI at flight segments M1 and M2. The figure
shows that the two plots are qualitatively different from each
other in terms of their dynamics within the manifold, which is
represented by their trajectories. This suggests that they have
different vector fields that govern their evolution within their
manifolds. Further, a thorough analysis indicates that the system
entropy (information content) at segment M2 is slightly higher
than that at segment M1. In other words, the heart IBI dynamics at
segment M2 appears slightly less structured compared to that at
segment M1. However, some quantitative evidence may be required to
support this observation, which has been done in the next section.
The above observations suggest that heart IBI dynamics depends on
the level of the pilot's perceived risk \cite{Sanjeev}.

\subsubsection{Approximate Entropy}

In general, entropy is concerned with information about the state
of a dynamical system. Approximate entropy ($ApEn$) is a useful
general measure of nonlinear dynamics. The technical details of
the $ApEn$ index have been provided in \cite{Pincus}. $ApEn$ is a
measure quantifying the regularity of the time series
\cite{Pikkujamsa}. $ApEn$ is defined as
\[
ApEn\left( {m,r} \right) = {\mathop {\lim} \limits_{N \to \infty}
} {\left[ {\phi \left( {m,r,N} \right) - \phi \left( {m + 1,r,N}
\right)} \right]}
\]
where $m$ is the maximum run length, $r$ is the tolerance, $N$ is
the total number of observations in the time series, and $\phi
\left( {m,r,N} \right)$ is the average value of the logarithm of
proportion of vectors that are closer than the $r$over all
vectors. $ApEn$ has a relatively lower value for a
structured/deterministic data set and a higher value for a
disordered/less-deterministic data set \cite{Ho}.

An algorithm has been provided in \cite{Kaplan} to compute $ApEn$.
The algorithm detects similar patterns in a time series and then
estimates the logarithmic likelihood that the following
observation will differ from the previously observed pattern.
Further, the pattern length and the pattern similarity criterion
can be manipulated while running the program \cite{Ho}. $ApEn$ can
be computed using no more than 100 observations \cite{Pincus}.

\cite{Sanjeev} employed $ApEn$ to analyze the heart IBI dynamics
of a pilot during flight segments M1 and M2. During the analysis,
the value for the maximum run length ($m$) was set equal to 2 and
the tolerance ($r$) was selected as 0.15 of the standard deviation
as suggested in \cite{Pincus}. The resulting $ApEn$ values at
segments M1 and M2 were 0.452 and 0.551 respectively. These
quantitative results indicate that the heart IBI $ApEn$ becomes
higher at segment M2 compared to that at segment M1. These results
re-confirm the visual analysis of the heart IBI Phase Plot data in
the previous section, where the entropy at segment M2 was believed
to be slightly higher than that at segment M1. This result may
further imply that the structure of the heart IBI data becomes
more disordered and unstructured during a higher level of
perceived-risk. Finally, these results suggest that the heart IBI
$ApEn$ may be a sensitive index of the level of perceived risk.

\subsection{Lyapunov Exponents in Time Series Data}

Recall that chaos arises from the exponential growth of
infinitesimal perturbations, together with global folding
mechanisms to guarantee boundedness of the solutions. This
exponential instability is characterized by the spectrum of
Lyapunov exponents\\ \cite{EckRuelle}. If one assumes a local
decomposition of the phase space into directions with different
stretching or contraction rates, then the spectrum of exponents is
the proper average of these local rates over the whole invariant
set, and thus consists of as many exponents as there are space
directions. The most prominent problem in time series analysis is
that the physical phase space is unknown, and that instead the
spectrum is computed in some embedding space. Thus the number of
exponents depends on the reconstruction, and might be larger than
in the physical phase space. Such additional exponents are called
\textsl{spurious}, and there are several suggestions to either
avoid them, or to identify them. Moreover, it is plausible that
only as many exponents can be determined from a time series as are
entering the Kaplan Yorke formula (see below). To give a simple
example: Consider motion of a high-dimensional system on a stable
limit cycle. The data cannot contain any information about the
stability of this orbit against perturbations, as long as they are
exactly on the limit cycle. For transients, the situation can be
different, but then data are not distributed according to an
invariant measure and the numerical values are thus difficult to
interpret. Apart from these difficulties, there is one relevant
positive feature: Lyapunov exponents are invariant under smooth
transformations and are thus independent of the measurement
function or the embedding procedure. They carry a dimension of an
inverse time and have to be normalized to the sampling interval
\cite{Tisean}.

\subsubsection{The Maximal Lyapunov Exponent}

The maximal Lyapunov exponent can be determined without the
explicit construction of a model for the time series. A reliable
characterization requires that the independence of embedding
parameters and the exponential law for the growth of distances are
checked \cite{Kantz,Rosenstein} explicitly. Consider the
representation of the time series data as a trajectory in the
embedding space, and assume that you observe a very close return
$s _{n^{\prime }}$ to a previously visited point $s _{n}$. Then
one can consider the distance \cite{Tisean}
\[
\Delta _{0}={s }_{n}-{s }_{n^{\prime }}
\]
as a small perturbation, which should grow exponentially in time.
Its future can be read from the time series:
\[
\Delta _{l}={s }_{n+l}-{s }_{n^{\prime }+l} .
\]
If one finds that $|\Delta _{l}|\approx \Delta _{0}e^{\lambda l}$ then $%
\lambda $ is (with probability one) the maximal Lyapunov exponent.
In practice, there will be fluctuations because of many effects,
which are discussed in detail in \cite{Kantz}. Based on this
understanding, one can derive a robust consistent and unbiased
estimator for the maximal Lyapunov exponent. One computes
\begin{equation}
S(\epsilon ,m,t)=\left\langle \ln \left(
\frac{1}{|\mathcal{U}_{n}|} \sum_{s _{n\prime }\in
\mathcal{U}_{n}}|s_{n+t}-s_{n\prime +t}|\right) \right\rangle
_{n}.  \label{eq:S}
\end{equation}
If $S(\epsilon ,m,t)$ exhibits a linear increase with identical
slope for all $m$ larger than some $m_{0}$ and for a reasonable
range of $\epsilon $, then this slope can be taken as an estimate
of the maximal exponent $\lambda _{1}$.

The formula is implemented using the Euclidean norm. Apart from
parameters characterizing the embedding, the initial neighborhood
size $\epsilon $ is of relevance: The smaller $\epsilon $, the
large the linear range of $S$, if
there is one. Obviously, noise and the finite number of data points limit $%
\epsilon $ from below. It is not always necessary to extend the average in (%
\ref{eq:S}) over the whole available data, reasonable averages can
be obtained already with a few hundred reference points $s_{n}$.
If some of the
reference points have very few neighbors, the corresponding inner sum in (%
\ref{eq:S}) is dominated by fluctuations. Therefore one may choose
to exclude those reference points which have less than, say, ten
neighbors. However, discretion has to be applied with this
parameter since it may introduce a bias against sparsely populated
regions. This could in theory affect the estimated exponents due
to multifractality. Like other quantities, Lyapunov estimates may
be affected by serial correlations
between reference points and neighbors. Therefore, a minimum time for $%
|n-n^{\prime }|$ can and should be specified here as well.
\begin{figure}[htb]
\centerline{\includegraphics[width=14cm]{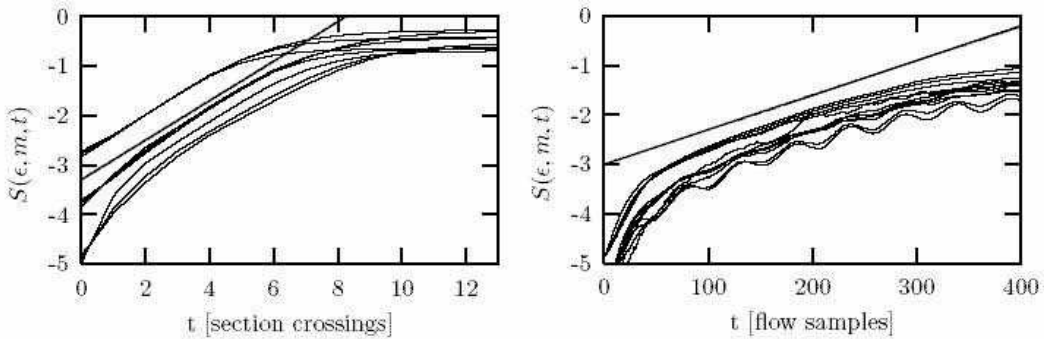}}
\caption{Estimating the maximal Lyapunov exponent of the CO2 laser
data (adapted from \cite{Tisean}). Left: results for the
Poincar\'e map data, where the average time interval $T_{av}$ is
52.2 samples of the flow, and the straight line indicates $\lambda
= 0.38$. Right: Lyapunov exponents determined directly from the
flow data. The straight line has slope $\lambda = 0.007$. In good
approximation, $\lambda_{map} = \lambda_{flow}T_{av}$. Here, the
time window w to suppress correlated neighbors has been set to
1000, and the delay time was 6 units.} \label{LyapTis}
\end{figure}

For example, the data underlying the top panel of Figure
\ref{LyapTis} are the values of the maxima of the CO$_{2}$ laser
data. Since this laser exhibits low dimensional chaos with a
reasonable noise level, we observe a clear linear increase in this
semi-logarithmic plot, reflecting the exponential divergence of
nearby trajectories. The exponent is $\lambda \approx 0.38$ per
iteration (map data!), or, when introducing the average time
interval, 0.007 per $\mu $s \cite{Tisean}.

\subsubsection{The Lyapunov Spectrum}

The computation of the full Lyapunov spectrum requires
considerably more effort than just the maximal exponent. An
essential ingredient is some estimate of the local Jacobians,
i.e.\ of the linearized dynamics, which rules the growth of
infinitesimal perturbations. One either finds it from direct fits
of local linear models of the type
\[
s_{n+1}=a_{n}s_{n}+b_{n},
\]
such that the first row of the Jacobian is the vector $a_{n}$, and
\[
(J)_{ij}=\delta _{i-1,j}\text{ \ \ \ \ \ for }i=2,\ldots ,m,
\]
where $m$ is the embedding dimension. The $a_{n}$ is given by the
least squares minimization
\[
\sigma ^{2}=\sum_{l}(s_{l+1}-a_{n}s_{l}-b_{n})^{2},
\]
where $\{s_{l}\}$ is the set of neighbors of $s_{n}$
\cite{Tisean}. Or one constructs a global nonlinear model and
computes its local Jacobians by taking derivatives. In both cases,
one multiplies the Jacobians one by one, following the trajectory,
to as many different vectors $u_{k}$ in tangent space as one wants
to compute Lyapunov exponents. Every few steps, one
applies a \textit{Gram--Schmidt orthonormalization procedure} to the set of $%
u_{k}$, and accumulates the logarithms of their rescaling factors.
Their average, in the order of the Gram-Schmidt procedure, give
the Lyapunov exponents in descending order. Apart from the problem
of spurious exponents, this method contains some other pitfalls:
It \textsl{assumes} that there exist well defined Jacobians, and
does not test for their relevance. In particular, when attractors
are thin in the embedding space, some (or all) of the local
Jacobians might be estimated very badly. Then the whole product
can suffer from these bad estimates and the exponents are
correspondingly wrong. Thus the global nonlinear approach can be
superior, if a modelling has been successful.

The computation of the first part of the Lyapunov spectrum allows
for some interesting cross-checks. It was conjectured
\cite{KaplanYorke}, and is found to be correct in most physical
situations, that the Lyapunov spectrum and the fractal dimension
of an attractor are closely related. If the expanding and least
contracting directions in space are continuously filled and only
one partial dimension is fractal, then one can ask for the
dimensionality of a (fractal) volume such that it is invariant,
i.e.\ such that the sum of the corresponding Lyapunov exponents
vanishes, where the last one is weighted with the non-integer part
of the \textit{Kaplan--Yorke dimension}:
\[
D_{KY}=k+\frac{\sum_{i=1}^{k}\lambda _{i}}{|\lambda _{k+1}|},
\]
where $k$ is the maximum integer such that the sum of the $k$
largest exponents is still non-negative. $D_{KY}$ is conjectured
to coincide with the information dimension.

The so--called \textit{Pesin identity} is valid under the same
assumptions and allows to compute the \textit{Kolmogorov--Sinai
entropy}:
\[
h_{KS}=\sum_{i=1}^{m}\Theta (\lambda _{i})\lambda _{i},
\]
where $\Theta$ is the Heaviside step function.

\subsection{Dimensions and Entropies of Time Series Data}

Solutions of dissipative dynamical systems cannot fill a volume of
the phase space, since dissipation is synonymous with a
contraction of volume elements under the action of the equations
of motion. Instead, trajectories are confined to lower dimensional
subsets which have measure zero in the phase space. These subsets
can be extremely complicated, and frequently they possess a
fractal structure, which means that they are in a nontrivial way
self-similar. Generalized dimensions are one class of quantities
to characterize this fractality. The \emph{Hausdorff dimension}
is, from the mathematical point of view, the most natural concept
to characterize fractal sets~\cite{EckRuelle}, whereas the
\emph{information dimension} takes into account the relative
visitation frequencies and is therefore more attractive for
physical systems. Finally, for the characterization of measured
data, other similar concepts, like the \emph{correlation
dimension}, are more useful. One general remark is highly relevant
in order to understand the limitations of any numerical approach:
dimensions characterize a set or an invariant measure whose
support is the set, whereas any data set contains only a finite
number of points representing the set or the measure. By
definition, the dimension of a finite set of points is zero. When
we determine the dimension of an attractor numerically, we
extrapolate from finite length scales, where the statistics we
apply is insensitive to the finiteness of the number of data, to
the infinitesimal scales, where the concept of dimensions is
defined. This extrapolation can fail for many reasons which will
be partly discussed below. Dimensions are invariant under smooth
transformations and thus again computable in time delay embedding
spaces.

Entropies are an information theoretical concept to characterize
the amount of information needed to predict the next measurement
with a certain precision. The most popular one is the
Kolmogorov-Sinai entropy. We will discuss here only the
correlation entropy, which can be computed in a much more robust
way. The occurrence of entropies in a section on dimensions has to
do with the fact that they can be determined both by the same
statistical tool \cite{Tisean}.

\subsubsection{Correlation Dimension}

Roughly speaking, the idea behind certain quantifiers of
dimensions is that the weight $p(\epsilon )$ of a typical
$\epsilon $-ball covering part of the invariant set scales with
its diameter like $p(\epsilon )\approx \epsilon ^{D}$, where the
value for $D$ depends also on the precise way one defines the
weight. Using the square of the probability $p_{i}$ to find a
point of
the set inside the ball, the dimension is called the correlation dimension $%
D_{2}$, which is computed most efficiently by the correlation sum
\cite{Tisean}
\[
C(m,\epsilon )={\frac{1}{N_{\mbox{\scriptsize pairs}}}}\sum_{j=m}^{N}%
\sum_{k<j-w}\Theta (\epsilon -|\text{s }_{j}-\text{s }_{k}|)\,,
\]
where $s _{i}$ are $m$-dimensional delay vectors, while
\[
N_{\mbox{\scriptsize pairs}}=(N-m+1)(N-m-w+1)/2
\]
is the number of pairs of points covered by the sums, $\Theta $ is
the Heaviside step function and $w$ will be discussed below. On
sufficiently small length scales and when the embedding dimension
$m$ exceeds the box-dimension of the attractor~\cite{SauerYorke},
\[
C(m,\epsilon )\propto \epsilon ^{D_{2}},
\]
Since one does not know the \textit{box-dimension} \textsl{a
priori}, one checks for convergence of the estimated values of
$D_{2}$ in $m$.

Fast implementation of the correlation sum have been proposed by
several
authors. At small length scales, the computation of pairs can be done in $%
O(N\log N)$ or even $O(N)$ time rather than $O(N^{2})$ without
loosing any of the precious pairs. However, for intermediate size
data sets we also need the correlation sum at intermediate length
scales where neighbor searching becomes expensive.

\subsubsection{Information Dimension}

Another way of attaching weight to $\epsilon $-balls, which is
more natural, is the probability $p_{i}$ itself. The resulting
scaling exponent is called the information dimension $D_{1}$.
Since the Kaplan--Yorke dimension is an approximation of $D_{1}$,
the computation of $D_{1}$ through scaling properties is a
relevant cross-check for highly deterministic data. $D_{1}$ can be
computed from a modified correlation sum, where, however,
unpleasant systematic errors occur. The \emph{fixed mass}
approach~\cite{badiipoliti} circumvents these problems, so that,
including finite sample corrections~\cite{grass_finite}, a rather
robust estimator exists. Instead of counting the number of points
in a ball one asks here for the diameter $\epsilon $ which a ball must have to contain a certain number $%
k$ of points when a time series of length $N$ is given. Its
scaling with $k$ and $N$ yields the dimension in the limit of
small length scales by
\[
D_{1}(m)=\lim_{k/N\rightarrow 0}{\frac{d\log
k/N}{d<\log\epsilon(k/N)>}}.
\]
Unlike the correlation sum, finite sample corrections are
necessary if $k$ is small. Essentially, the $\log $ of $k$ has to
be replaced by the digamma function $\Psi (k)$. Given $m$ and
$\tau $, the routine varies $k$ and $N$ such that the largest
reasonable range of $k/N$ is covered with moderate computational
effort. This means that for $1/N\leq k/N\leq K/N$ (default:
$K=100$), all $N$ available points are searched for neighbors and
$k$ is varied. For $K/N<k/N\leq 1$, $k=K$ is kept fixed and $N$ is
decreased \cite{Tisean}.

\subsubsection{Entropy estimates}

The correlation dimension characterizes the $\epsilon $ dependence
of the correlation sum inside the scaling range. It is natural to
ask what we can learn form its $m$-dependence, once $m$ is larger
than $D_{0}$. The number of $\epsilon $-neighbors of a delay
vector is an estimate of the local
probability density, and in fact it is a kind of joint probability: All $m$%
-components of the neighbor have to be similar to those of the
actual vector simultaneously. Thus when increasing $m$, joint
probabilities covering larger time spans get involved. The scaling
of these joint probabilities is related to the correlation entropy
$h_{2}$, such that
\[
C(m,\epsilon )\approx \epsilon ^{D_{2}}e^{-mh_{2}},
\]
As for the scaling in $\epsilon $, also the dependence on $m$ is
valid only asymptotically for large $m$, which one will not reach
due to the lack of data points. So one will study $h_{2}(m)$
versus $m$ and try to extrapolate to large $m$. The correlation
entropy is a lower bound of the Kolmogorov--Sinai entropy, which
in turn can be estimated by the sum of the positive Lyapunov
exponents.

An alternate means of obtaining these and the other generalized
entropies is by a box counting approach. Let $p_{i}$ be the
probability to find the system state in box $i$, then the order
$q$ entropy is defined by the limit of small box size and large
$m$ of
\[
\sum_{i}p_{i}^{q}\approx e^{-mh_{q}}.
\]
To evaluate $\sum_{i}p_{i}^{q}$ over a fine mesh of boxes in
$m>>1$ dimensions, economical use of memory is necessary: A simple
histogram would take $(1/\epsilon )^{m}$ storage \cite{Tisean}.


\begin{thebibliography} {99}
\bibitem[Berkowitz \& Papiernik, 1993]{Berkowitz} Berkowitz, G.S., Papiernik, E.: Epidemiology of
Preterm Birth. Epidem. Rev. \textbf{15}(2), 414--443, (1993)

\bibitem[Robillard {\it et al.}, 2007]{Dekker1} Robillard, P.-Y.,
Dekker, G., Chaouat, G., Hulsey, T.C.: Etiology of preeclampsia:
maternal vascular predisposition and couple disease—mutual
exclusion or complementarity? J. Reprod. Immun. \textbf{76}, 1–-7,
(2007)

\bibitem[Rush {\it et al.}, 1976]{Rush} Rush, R.W., Keirse, M.J., Howat, P., et al.: Contribution
of preterm delivery to perinatal mortality. Br. Med. J. \textbf{2}, 965-8, (1976)

\bibitem[Overpeck {\it et al.}, 1992]{Overpeck} Overpeck, M.D., Hoffman, H.J., Prager, K.: The
lowest birth-weight infants and the US infant mortality rate: NCHS 1983 linked birth/infant death. Am. J. Public Health \textbf{82}, 441--444, (1992)

\bibitem[Nat. Acad. Sci, 1985]{Inst} Institute of Medicine, National Academy of
Sciences: Preventing low birth weight. Washington, DC: Nat. Acad. Press, (1985)

\bibitem[Arias \& Tomich, 1993]{Arias} Arias, F., Tomich, P.: Etiology and outcome of
low birth weight and preterm infants. Obstet. Gynecol. \textbf{60}, 277-81, (1982)

\bibitem[Kitchen {\it et al.}, 1992]{Kitchen} Kitchen, W.F., Doyle, L.W., Ford, E.G., et al.: Very
low birth weight and growth to age 8 years. I. Weight and height. Am. J. Dis. Child. \textbf{146}, 40--5, (1992)

\bibitem[Ross {\it et al.}, 1990]{Ross} Ross, G., Upper, E.G., Auld, P.A.M.: Growth
achievement of very low birth weight premature children at school age. J. Pediatr. 117, 307--309, (1990)

\bibitem[Escobar {\it et al.}, 1991]{Escobar} Escobar, G.J., Littenberg, B., Petitti, D.B.: Outcome
among surviving very low birth weight infants: a meta-analysis. Arch. Dis. Child. \textbf{66}, 204-211, (1991)

\bibitem[Veen {\it et al.}, 1991]{Veen} Veen, S., Ens-Dokkum, M.H., Schreuder, A.M., et al.: Impairments, disabilities, and handicaps of
very preterm and very-low-birthweight infants at five years of age: The Collaborative Project on Preterm and Small for Gestational Age Infants (POPS) in the Netherlands. Lancet, \textbf{338}, 33-336, (1991)

\bibitem[Kessel {\it et al.}, 1984]{Kessel} Kessel, S.S., Villar, J., Berendes, H.W., et al.: The
changing pattern of low birth weight in the United States—1970 to 1980. JAMA, \textbf{251}, 1978-82, (1984)

\bibitem[Kleinman \& Kessel, 1981]{Kleinman} Kleinman, J.C., Kessel, S.S.: The recent decline in
infant mortality: Health, United States 1980. Washington, DC: US Department of Health, Education, and Welfare. DHEW publication no. 81-1232, (1981)

\bibitem[World Health Org. 1977]{WHT1} World Health Organization. International classification
of diseases: manual of the international statistical classification of diseases, injuries,
and causes of death. Ninth Revision. Geneva, Switzerland: World Health Organization,
(1977)

\bibitem[Oczeretko {\it et al.}, 2006]{Oczeretko1} Oczeretko, E., Swiatecka, J., Kitlas,
A., Laudanski, T., Pierzynski, P.: Visualization of
synchronization of the uterine contraction signals: Running
cross-correlation and wavelet running cross-correlation methods.
Med. Eng. Phys. \textbf{28}, 75–81, (2006)

\bibitem[Radharkrishnan {\it et al.}, 2000]{Radharkrishnan} Radharkrishnan, N., Wilson, J.D., Lowery, C., Eswaran, H.,
Murphy, P.: A fast algorithm for detecting contractions in uterine
electromyography. IEEE Eng. Med. Biol. \textbf{19}(2), 89–94,
(2000)

\bibitem[Pierzynski {\it et al.}, 2007]{Pierzynski} Pierzynski, P., Oczeretko, E., Laudanski, P., Laudanski, T.: New research models and novel signal analysis in studies on
preterm labor: a key to progress? BMC Pregn. Childbirth, \textbf{7}(Suppl 1), S6, (2007)

\bibitem[Oczeretko {\it et al.}, 2005]{Oczeretko} Oczeretko, E., Kitlas, A., Swiatecka, J., Borowska, M., Laudanski, T.: Nonlinear
dynamics in uterine contractions analysis. In Fractals in Biology and Medicine, Volume IV. Edited by: Losa G, Merlini D,
Nonnemacher T, Weibel E. Birkh\"auser Verlag, Basel, 215-222, (2005)

\bibitem[Oczeretko {\it et al.}, 2004]{Oczeretko0} Oczeretko, E., Kitlas, A., Swiatecka, J., Laudanski, T.: Fractal analysis
of the uterine contractions. Rivista di Biologia/Biology Forum,
\textbf{97}(3), 499-504, (2004)

\bibitem[Badii \& Politi, 1985]{badiipoliti} Badii, R., Politi, A.: Statistical description of chaotic attractors. J. Stat. Phys. {\bf 40}, 725, (1985)

\bibitem[Beck \& Schl\"ogl, 1993]{BS93} Beck, C., Schl\"ogl, F.: Thermodynamics of chaotic systems. Cambridge Univ. Press, Cambridge, (1993)

\bibitem[Benettin {\it et al.}, 1980]{BGGS80} Benettin, G., Giorgilli, A., Galgani, L., Strelcyn, J.M.: Lyapunov exponents for smooth dynamical systems and for Hamiltonian systems; a method for computing all of them. Part 1: theory, and Part 2: numerical applications. Meccanica, \textbf{15}, 9-20 and 21-30, (1980)

\bibitem[Benettin, 1984]{B84} Benettin, G.: Power law behaviour of Lyapunov exponents in some conservative dynamical systems. Physica D \textbf{13}, 211-213, (1984)

\bibitem[Boffetta {\it et al.}, 2001]{BLV01} Boffetta, G., Lacorata, G., Vulpiani, A.: Introduction to chaos and diffusion. Chaos in geophysical flows, ISSAOS, (2001)

\bibitem[Boffetta {\it et al.}, 2002]{PredictCompl} Boffetta, G., Cencini, M., Falcioni, M., Vulpiani, A.: Predictability: a way to characterize Complexity, Phys. Rep., \textbf{356}, 367--474, (2002)

\bibitem[Caiani \emph{et al.}, 1997]{CCC97} L. Caiani, L. Casetti, C. Clementi, M. Pettini, Geometry of Dynamics, Lyapunov Exponents, and Phase Transitions. Phys. Rev. Lett. \textbf{79}, 4361--4364, (1997)

\bibitem[Chen \& Dong, 1998]{ChOrder} Chen, G., Dong, X.: From Chaos to Order: Methodologies, Perspectives and Application. Singapore: World Scientific, 1998.

\bibitem[Chirikov, 1979]{chirikov79} Chirikov, B.V.: A universal instability of many-dimensional oscillator systems. Phys. Rep. \textbf{52}, 264-379, (1979)

\bibitem[Eckmann \& Ruelle, 1985]{EckRuelle} Eckmann, J.P., Ruelle, D.: Ergodic theory of chaos and strange attractors. Rev. Mod. Phys. {\bf 57}, 617, (1985)

\bibitem[Eisenhart, 1929]{Eisenhart} Eisenhart, L.P.: Dynamical trajectories and geodesics. Math. Ann. \textbf{30}, 591-606, (1929)

\bibitem[Franzosi \& Pettini, 2004]{FP04} Franzosi, R., Pettini, M.: Theorem on the origin of Phase Transitions. Phys. Rev. Lett. \textbf{92}(6), 060601, (2004)

\bibitem[Franzosi \emph{et al.}, 1999]{FCS99} R. Franzosi, L. Casetti, L. Spinelli, M. Pettini, Topological aspects of geometrical signatures of phase transitions. Phys. Rev. E \textbf{60}, 5009, (1999)

\bibitem[Franzosi \emph{et al.}, 2000]{FPS00} R. Franzosi, M. Pettini, L. Spinelli, Topology and phase transitions: a paradigmatic evidence. Phys. Rev. Lett. \textbf{84}, 2774--2777, (2000)

\bibitem[Grassberger, 1988]{grass_finite} Grassberger, P.: Finite sample corrections to entropy and dimension estimates. Phys. Lett. A {\bf 128}, 369, (1988)

\bibitem[Grote \& Sch\"{o}ner, 2006]{GS} Grote, C., Sch\"{o}ner, G.: Context-sensitive generation of goal-directed behavioral sequences based on neural attractor dynamics. Proceeedings of the ISR/ROBOTIK2006 Joint Conference on Robotics, Munich, Germany, May, (2006)

\bibitem[Haken, 1983]{Haken} Haken, H.: Synergetics: An Introduction (3rd ed). Springer, Berlin, (1983)

\bibitem[Haken, 1993]{Haken2} Haken, H.: Advanced Synergetics: Instability Hierarchies of Self-Organizing Systems and Devices (3nd ed). Springer, Berlin, (1993)

\bibitem[Haken, 2002]{HakenBrain} Haken, H.: Brain Dynamics, Synchronization and Activity Patterns in Pulse-Coded Neural Nets with Delays and Noise, Springer, New York, (2002)

\bibitem[Hegger {\it et al.}, 1999]{Tisean} Hegger, R., Kantz, H., Schreiber, T.: Practical implementation of nonlinear time series methods: The TISEAN package. CHAOS \textbf{9}, 413 (1999)

\bibitem[Hilborn, 1994]{Hilborn} Hilborn, R.C.: Chaos and Nonlinear Dynamics: An Introduction for Scientists and Engineers. New York: Oxford University Press, (1994)

\bibitem[Ho {\it et al.}, 1997]{Ho} Ho, K.K.L. \emph{et al.}: Predicting survival in heart failure case control subjects by use of fully automated methods for deriving nonlinear and conventional indices of heart rate dynamics.  Circulation, \textbf{96},  842-848, {1997}

\bibitem[Hodgkin \& Huxley, 1952]{H-H} Hodgkin, A.L., Huxley, A.F.: A quantitative description of membrane current and application to conduction and excitation in nerve. J. Physiol., \textbf{117}, 500--544, (1952)

\bibitem[Hoppensteadt \& Izhikevich, 1997]{Hop-Iz} Hoppensteadt, F.C., Izhikevich, E.M.: Weakly Connected Neural Networks. Springer, New York, (1997)

\bibitem[Ivancevic \& Ivancevic, 2006a]{StrAttr} Ivancevic, V., Ivancevic, T.: High-Dimensional Chaotic and Attractor Systems. Springer, Berlin, (2006)

\bibitem[Ivancevic \& Ivancevic, 2006b]{GaneshSprBig} Ivancevic, V., Ivancevic, T.: Geometrical Dynamics of Complex Systems. Springer, Series: Microprocessor-Based and Intelligent Systems Engineering, \textbf{31}, Dordrecht, (2006)

\bibitem[Ivancevic \& Ivancevic, 2006c]{NatBiodyn} Ivancevic, V., Ivancevic, T.: Natural Biodynamics. World Scientific, Series: Mathematics, (2006)

\bibitem[Ivancevic \& Ivancevic, 2007a]{NeuFuz} Ivancevic, V., Ivancevic, T.: Neuro-Fuzzy Associative Machinery for Comprehensive Brain and Cognition Modelling. Springer, Berlin, (2007)

\bibitem[Ivancevic \& Ivancevic, 2007b]{GaneshADG} Ivancevic, V., Ivancevic, T.: Applied Differfential Geometry: A Modern Introduction. World Scientific, Series: Mathematics, (2007)

\bibitem[Izhikevich, 1999]{IzhikevichClass} Izhikevich, E.M.: Class 1 neural excitability, conventional synapses, weakly connected networks, and mathematical foundations of pulse-coupled models. IEEE Trans. Neu. Net., \textbf{10}, 499--507, (1999)

\bibitem[Izhikevich, 2001]{IzhikevichResn} Izhikevich, E.M.: Resonate-and-fire neurons. Neu. Net., \textbf{14}, 883--894, (2001)

\bibitem[Izhikevich, 2004]{Izhikevich1} Izhikevich, E.M.: Which model to use for cortical spiking neurons? IEEE Trans. Neu. Net., \textbf{15}, 1063--1070, (2004)

\bibitem[Kantz, 1994]{Kantz} Kantz, H.: A robust method to estimate the maximal Lyapunov exponent of a time series. Phys. Lett. A \textbf{185}, 77, (1994)

\bibitem[Kaplan \& Yorke, 1987]{KaplanYorke} Kaplan, J., Yorke, J.: Chaotic behavior of multidimensional difference equations. In Peitgen, H.~O. \& Walther, H.~O. (eds.) Functional Differential Equations and Approximation of Fixed Points. Springer, New York, (1987)

\bibitem[Kaplan {\it et al.}, 1991]{Kaplan} Kaplan, D.T. \emph{et al.}: Aging and the complexity of cardiovascular dynamics. Biophys. J., \textbf{59}, 945-949, (1991)

\bibitem[Kuramoto, 1984]{Kuramoto} Kuramoto, Y.: Chemical Oscillations. Waves and Turbulence. Springer, New York, (1984)

\bibitem[Lorenz, 1963]{Lorenz} Lorenz, E.N.: Deterministic Nonperiodic Flow. J. Atmos. Sci., \textbf{20}, 130--141, (1963)

\bibitem[Mattfeldt, 1997]{Mattfeldt} Mattfeldt, T.: Nonlinear deterministic analysis of tissue texture: a stereological study on mastopatic and mammary cancer tissue using chaos theory. J. Microscopy, 185(1), 47--66, (1997)

\bibitem[Morris \& Lecar, 1981]{Morris} Morris, C., Lecar, H.: Voltage oscillations in the barnacle giant muscle fiber. Biophys. J., \textbf{35}, 193--213, (1981)

\bibitem[Nagumo {\it et.~al.}, 1960]{Nagumo} Nagumo, J., Arimoto, S., Yoshizawa, S.: An active pulse transmission line simulating 1214-nerve axons, Proc. IRL, \textbf{50}, 2061--2070, (1960)

\bibitem[Ott {\it et al.}, 1990]{OGY} Ott, E., Grebogi, C., Yorke, J.A.: Controlling chaos. Phys. Rev. Lett., \textbf{64}, 1196--1199, (1990)

\bibitem[Ott, 1993]{Ott93} Ott, E.: Chaos in dynamical systems. Cambridge University Press, Cambridge, (1993)

\bibitem[Pikkujamsa {\it et al.}, 1999]{Pikkujamsa} Pikkujamsa, S.M. \emph{et al.}: Cardiac interbeat interval dynamics from childhood to senescence: Comparison of conventional and new measures based on fractals and chaos. Circulation, \textbf{100}, 393-399, (1999)

\bibitem[Pincus, 1995]{Pincus} Pincus, S.: Approximate entropy (ApEn) as a complexity measure. Chaos, \textbf{5}, 110-117, (1995)

\bibitem[Rose \& Hindmarsh, 1989]{Rose} Rose, R.M., Hindmarsh, J.L.: The assembly of ionic currents in a thalamic neuron. I The three-dimensional model. Proc. R. Soc. Lond. B, \textbf{237}, 267--288, (1989)

\bibitem[Rosenstein {\it et.~al}, 1993]{Rosenstein} Rosenstein, M.T., Collins, J. J., De Luca, C.J.: A practical method for calculating largest Lyapunov exponents from small data sets. Physica D \textbf{65}, 117, (1993)

\bibitem[Sauer \& Yorke, 1993]{SauerYorke} Sauer, T. Yorke, J.: How many delay coordinates do you need? Int. J. Bifur. Chaos {\bf 3}, 737, (1993)

\bibitem[Sharma, 2006]{Sanjeev} Sharma, S.: An Exploratory Study of Chaos in Human-Machine System Dynamics, IEEE Trans. SMC B. 36(2), 319-326, (2006)

\bibitem[Sparrow, 1982]{Sp} Sparrow, C.: The Lorenz Equations: Bifurcations, Chaos, and Strange Attractors. Springer, New York, (1982)

\bibitem[Williams, 1997]{Williams} Williams, G.P.: Chaos Theory Tamed. Joseph Henry, Washington, D.C. (1997)

\end{thebibliography}
\end{document}